\PassOptionsToPackage{square,sort,comma,numbers}{natbib}

\documentclass{article}  

\usepackage[preprint]{nips_2018}

\def\BibTeX{{\rm B\kern-.05em{\sc i\kern-.025em b}\kern-.08emT\kern-.1667em\lower.7ex\hbox{E}\kern-.125emX}}

\usepackage{booktabs} 

\usepackage{array}
\usepackage{breakcites}
\usepackage{cuted}
\usepackage{capt-of}
\usepackage[font=small,labelfont=bf,tableposition=top]{caption}
\usepackage{psfrag}
\usepackage{color}
\usepackage{balance}
\usepackage{multirow}
\usepackage{amsthm, amsmath, amssymb, amsfonts}
\usepackage{graphicx}
\usepackage{tabularx}

\begin{document}

\title{Modality Dropout for Improved Performance-driven Talking Faces}

\author{Ahmed Hussen Abdelaziz\thanks{Both authors contributed equally to this paper.}\\
Apple\\
Cupertino, CA\\
\texttt{ahussenabdelaziz@apple.com}
\And
Barry-John Theobald\footnotemark[1]\\
Apple\\
Cupertino, CA\\
\texttt{bjtheobald@apple.com}
\And
Paul Dixon\\
Apple\\
Zurich\\
\texttt{dixonp@apple.com}
\AND
Reinhard Knothe\\
Apple\\
Zurich\\
\texttt{knothe@apple.com}
\And
Nicholas Apostoloff\\
Apple\\
Cupertino, CA\\
\texttt{napostoloff@apple.com}
\And
Sachin Kajareker\\
Apple\\
Cupertino, CA\\
\texttt{skajarekar@apple.com}
}

\maketitle

\begin{abstract}
We describe our novel deep learning approach for driving animated faces using both acoustic and visual information.  In particular, speech-related facial movements are generated using audiovisual information, and non-speech facial movements are generated using only visual information.  To ensure that our model exploits both modalities during training, batches are generated that contain audio-only, video-only, and audiovisual input features.  The probability of dropping a modality allows control over the degree to which the model exploits audio and visual information during training.  Our trained model runs in real-time on resource limited hardware (e.g.\ a smart phone),  it is user agnostic, and it is not dependent on a potentially error-prone transcription of the speech.  We use subjective testing to demonstrate: 1) the improvement of audiovisual-driven animation over the equivalent video-only approach, and 2) the improvement in the animation of speech-related facial movements after introducing modality dropout.  Before introducing dropout, viewers prefer audiovisual-driven animation in 51\% of the test sequences compared with only 18\% for video-driven.  After introducing dropout viewer preference for audiovisual-driven animation increases to 74\%, but decreases to 8\% for video-only.
\end{abstract} 

\vspace{2mm}

\section{Introduction} \label{sec:intro}

Performance-driven facial animation involves re-targeting facial motion from an actor to a character model via animation controls.  Traditionally, motion is captured using a marker-based system \cite{williams1990performance}, an RGB-D camera \cite{Bouaziz, Li2013, weise2011realtime}, a stereo-camera pair \cite{beeler2011high}, or a monocular camera \cite{Bolkart2015AGM, Cao2015, Thies2018}.  A challenge is capturing and transferring high fidelity motion to ensure the expressiveness of the actor is honored.  Furthermore, it is important to preserve perceptually important details in the facial motion, e.g.\ lip closures, which can be subtle and are susceptible to being masked by tracking noise.  For these reasons, dense facial features might be required, which results in a trade-off between global and local fidelity.

To avoid the need for run-time tracking of dense features to capture high-fidelity lip motion, animation controls have been learned from spectral features extracted from speech \cite{Karras2017, HussenAbdelaziz2019, thies2019nvp, Daniel19, Najmeh18} and from lexical features extracted from phonemes \cite{Wang2008ART, Parker2017ExpressiveVT, Latif2017VariationalAF, taylor2017deep}.  However, a limitation is that animation controls are learned only for speech-related facial motion, and non-speech-related information, e.g.\ blinking, must be learned by other means.

\begin{figure}[h]
\centering
\includegraphics[width=0.8\linewidth]{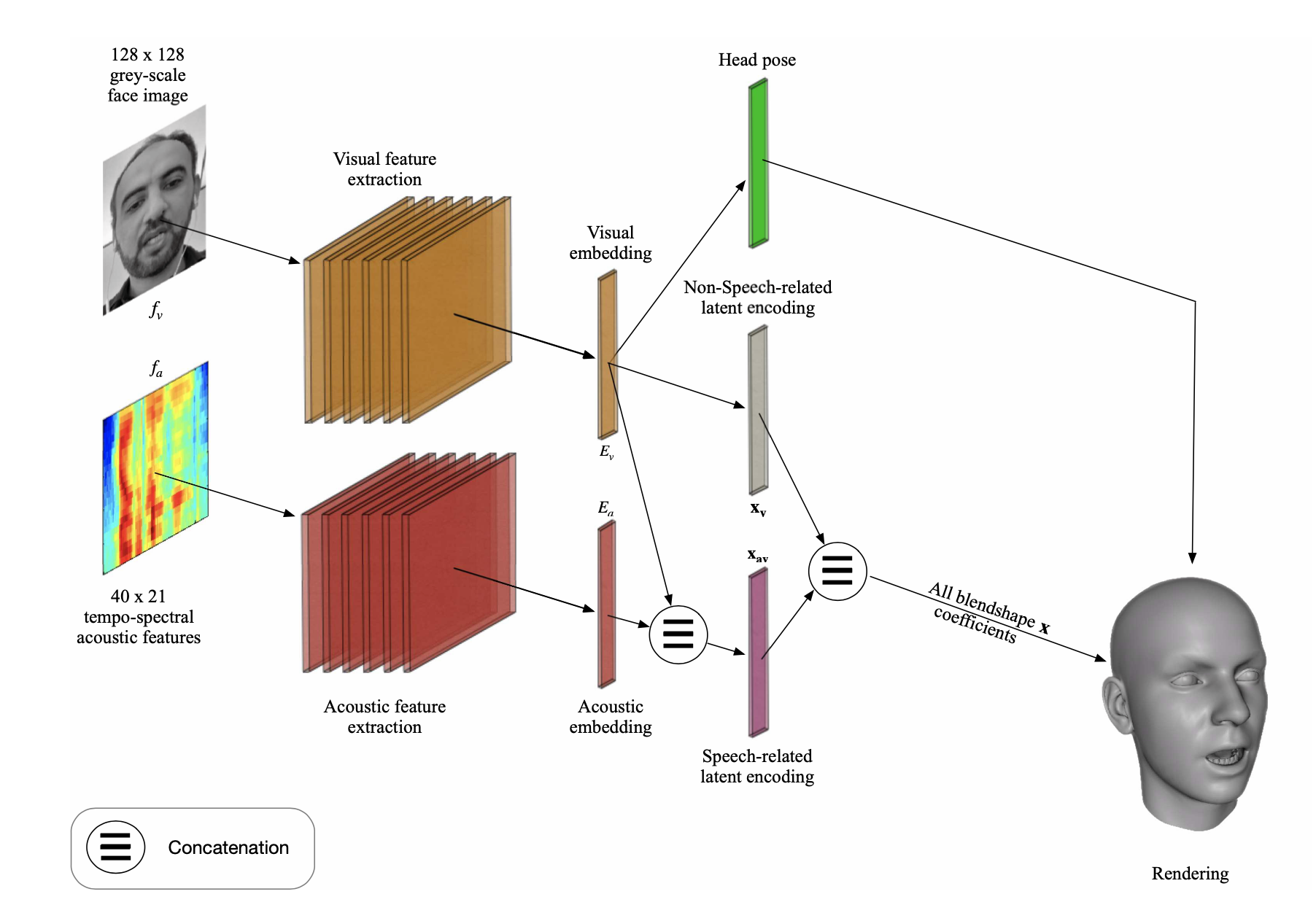}
\caption{Neural network architecture for audiovisual performance-driven 3D talking heads. Audio and video embeddings are fused to generate blendshape coefficients for driving the mouth and jaw, while only video embeddings are used to generate the blendshape coefficients for the rest of the face and the head pose.  See Section \ref{sec:avft} for details.}
\label{fig:framework}
\end{figure}

In this paper, we introduce a neural network that jointly uses audio and video features to improve performance-driven animation of 3D talking faces.  Our approach faithfully captures facial expressions and head pose from visual tracking in images and improves the animation of speech-related lip movements by incorporating features extracted from acoustic speech.   The audio and video inputs are processed by stacks of convolutional layers to extract embeddings, which are fused by concatenating and feeding into an affine layer to regress to speech-related latent vectors.  Non-speech-related latent vectors are estimated using a separate affine layer that is applied only to the video embeddings. The speech- and non-speech-related latent vectors are cascaded and mapped to the full set of facial controls.  A separate affine layer is used to estimate the head pose from the video embeddings. Finally, the generated facial controls and head pose are used to render a 3D mesh of the talking face --- see Figure \ref{fig:framework} for an overview.

Training a neural network for performance-driven facial capture that uses both audio and visual inputs is challenging,  The facial movements in the visual modality are strongly correlated with the output animation controls, and the network ignores information in the acoustic modality during training.  To avoid this we use an approach similar to \emph{moddrop} \cite{Neverova2015}, which was originally proposed to improve the robustness of a multimodal gesture recognition network to missing inputs by learning cross-modality correlations.  Here we show that by reducing the correlation between the inputs and the output, the network is forced to use information from both input modalities, which results in better quality capture of lip-motion in  performance-driven facial capture.  We demonstrate the efficacy of our approach using subjective assessment to measure the improvement of our audiovisual network over a video-only baseline, and the further gain achieved after introducing modality dropout.  Our main contributions are: 
\begin{itemize}
\item{the introduction of an audiovisual network architecture that extracts and fuses embeddings from audio and video inputs to generate high-quality performance-driven 3D facial animation independently of the speaker,}
\item{demonstration and evaluation of modality dropout during training to force the network to use information from both modalities, even when one is so highly correlated with output that the other would otherwise largely be ignored,}
\item{subjective evaluations comparing video-only and audiovisual systems, the effect of introducing and varying the amount of modality dropout used during training, and the effect of including future acoustic frames as context in the input for a non-real-time system.}
\end{itemize}

The remainder of this paper is structured as follows: In Section~\ref{sec:rw}, we discuss related works. Section~\ref{sec:bsc} describes the problem of the offline extraction of facial controls from video sequences. Section~\ref{sec:avft} outlines our model architectures and describes the modality dropout training strategy. The experiments used to evaluate the model performance and the results are described in Section~\ref{sec:exp}. Finally, we conclude the paper and give an outline of future work in Section~\ref{sec:conclusion}.

\section{Related work} \label{sec:rw}

Prior work on facial performance capture has traditionally used video-only methods or audio-only methods.  Video-based methods re-target a visual representation of the motion of the face directly to a model, whereas audio-only methods use regression to predict facial motion from acoustic speech.  An advantage of video-based approaches is the motion is observed, but a limitation is that not all facial motion is equally important perceptually.  Thus optimizing for the overall best performance may not result in the most pleasing output.  An advantage of audio-only methods is that the re-targeting can learn to exploit the correlation between audio and visual information, but not all facial motion can be predicted from acoustics.  A compromise would be to use both modalities and learn to fuse the pertinent audio and visual information for reliable performance capture.

\textbf{Video-driven performance-based facial animation} most commonly involves tracking facial features in image sequences from either an RGB or an RGB-D camera. The motion of the features is then transferred to either a user-dependent rig \cite{Alexander2009, garrido:2015:vdub, Wu2016, Laine2017} or a personalization of a generic face model \cite{Cao2014, Cao2015, Bolkart2015AGM}.  To improve the robustness of the facial feature tracking, and thus the quality of the retargeting, it is common to first train a model offline to provide constraints on the facial feature tracking using either a user-specific blendshape model \cite{weise2011realtime, Bouaziz, Li2013}, or a statistical model in the form of multi-linear models \cite{Bolkart2015AGM}.  In this paper, we use a method similar to that in \cite{weise2011realtime} for offline tracking.  However, we extend the computation of the per-frame blendshape coefficients by adding acoustic information to provide additional priors on the blendshape coefficients.

\textbf{Audio-driven facial animation} approaches can be broadly classified as being either:  (1) direct, which uses low-level acoustic features extracted from the speech signal, or (2) indirect, which uses an abstract representation of the speech, e.g.\ a phonemic transcription from an automatic speech recognizer.  For the direct approaches, the audio-to-animation conversion function mostly uses some form of regression \cite{Daniel19, Karras2017, Liu2015, pham2017end, shimba2015talking, song2018talking, suwaj2017, taylor2016audio, taylor2017deep, HussenAbdelaziz2019} or codebook indexing using acoustic features extracted from the speech \cite{arslan1998codebook,gutierrez2005speech}.  For indirect approaches, the mapping function involves concatenation or interpolation of pre-existing data \cite{bregler1997video,cosatto2000photo,ezzat2002trainable,taylor2012dynamic,mattheyses2013comprehensive} or using a generative model \cite{anderson2013expressive,fan2015photo,jalalifar2018speech,kim2015decision,vougioukas2018end}.

In this work we simplify the network architecture and make it real-time-capable by using standard Mel-scaled filter-bank audio features.  We use the visual modality to make the overall network speaker- and language-independent, and increase the robustness to cross-speaker distortion and acoustic noise.

\textbf{For Audio-visual-based facial animation} models, coupled hidden Markov models (CHMMs) \cite{brand1999voice} can be used. CHMMs model the asynchrony between speech sounds and lip shapes explicitly through cross-time and cross-chain conditional probabilities \cite{xie2007coupled,abdelazizCHMM}. For decoding a state sequence, from which visual parameters are sampled, \cite{choi2001hidden,fu2005audio} used Baum--Welch HMM inversion instead of the commonly used Viterbi decoding, which results in more accurate animation controls,

HMMs allow for only a single hidden state to be occupied in each time frame. This limitation means that many states are required to model multimodal signals than would otherwise be necessary to capture the complexities of the cross-modal dynamics. \cite{xie2007realistic} overcomes this by using dynamic Bayesian networks (DBNs) with Baum--Welch DBN inversion to model the cross-model dependencies and to generate animation from speech.

A  unit-selection-based system was introduced in \cite{Liu2015}, where dynamic programming is applied to choose from a pre-collected audio-visual database a candidate sequence for each input frame. The selection of the candidate frames is computed based on a weighted sum of distances between inferred audio and visual frames and candidate frames from the database. The weights are computed based on hand-crafted reliability measures of the audio and video streams. 

Compared to all audio-visual approaches above, we train an audio-visual network end-to-end, where the stream reliability measures are computed as parts of the fusion layer in the network.

\textbf{Fusion schemes for audio-visual neural networks} have been investigated in applications, such as sentiment analysis \cite{Amir17},  emotion recognition \cite{Song2004}, speech recognition \cite{Afouras18},  gesture recognition \cite{Neverova2015}, voice activity detection \cite{soo18}, speaker verification \cite{Suwon19}, and speech enhancement \cite{Ariel18}. The simplest scheme for fusion in all of these applications is \emph{direct fusion} \cite{abdelaziz2017comparing}, also known as early fusion. In this fusion scheme, the raw audio and video modalities are cascaded and then fused using a stack of nonlinear layers. Rather than cascading the raw data directly, the audio and video streams can first be pre-processed by stacks of hidden layers before feeding them into fusion layers \cite{Ariel18}.  In multimodal fusion, it is useful to consider the reliability of each stream, and then weight the streams accordingly.  In direct fusion the reliability of each stream is considered implicitly.  However, other schemes, such as gated fusion \cite{Ovalle2017GatedMU}, model the reliability of each stream explicitly.  Other fusion schemes include Tensor fusion \cite{Amir17}, which accounts for intra-modality and inter-modality dynamics, and the use of attention mechanisms for audio-visual sequence-to-sequence models \cite{Chung2016LipRS} . 

In this paper, we use an early fusion scheme with a single affine fusion layer.  We have found that this fusion coupled with modality dropout during training is sufficient to produce good quality facial animation in real-time, and on resource constrained hardware.

\section{Extracting blendshape coefficients}\label{sec:bsc}

The space of facial motion is represented using a generic blendshape model inspired by the Facial Action Coding System (FACS)~\cite{Ekman1978}.  Using this model, the vertices of a 3D mesh corresponding to a facial expression are given by:
\begin{equation}
\mathbf{v}(\mathbf{x}) = \mathbf{b}_{0} + \mathbf{B}\mathbf{x},
\end{equation}
where $\mathbf{v}(\mathbf{x})$ are the mesh vertices, $\mathbf{b}_0$ is the neutral position of the vertices, the columns of $\mathbf{B}$ define additive deviations from $\mathbf{b}_0$ (blendshapes), and $\mathbf{x} \in {[0,1]}$ define that contribution of each blendshape in the representation of $\mathbf{v}(\mathbf{x})$.

The goal of performance-driven facial animation is to transfer the motion of the face of an actor to a model via the corresponding blendshape coefficients.  We treat this as a regression task and train a DNN to map from input face images to blendshape coefficients.  However, labelling these coefficients manually to train a DNN is prohibitively time consuming, so we estimate blendshape coefficients and head pose using an extension of the method in~\cite{weise2011realtime}.

To generate blendshape coefficients for training the DNN, we construct a personalized model for each subject in our dataset by adapting a generic blendshape model using example-based facial rigging~\cite{Li2010} with added positional constraints from 2D landmarks~\cite{He2017AFE}.  In particular, we use an RGB-D camera to record each subject maintaining a series of prototypical facial expressions, including neutral, while rotating their head.  The personalized model is then created by modifying the generic blendshape model to best match each facial expression.

The personalized model for a user is then used to generate the head motion labels and blendshape coefficients for all video frames for that user by first rigidly aligning the model to the depth maps using iterative closest point (ICP) with point-plane constraints, and then solving for the blendshape coefficients that best explain the input data.  The optimization function is composed of two losses:  The first is a point-to-plane fitting loss on the depth maps:
\begin{equation}
{D_i}(\mathbf{x}) ={  \left( {\mathbf{n}_i}^\top  \left( \mathbf{v}_i(\mathbf{x}) - \overline{\mathbf{v}}_i \right)  \right) }^{2},
\label{eq2}
\end{equation}
where $\mathbf{v}_i(\mathbf{x})$ is the $i^{th}$ vertex displacement of the mesh as a function of the blendshape coefficients, $\overline{\mathbf{v}}_i$ is the projection of $\mathbf{v}_i$ onto the depth map, and $\mathbf{n}_i$ is surface normal of $\overline{\mathbf{v}}_i$.  The second loss is a fitting loss on the 2D landmarks:
\begin{equation}
{L_j}(\mathbf{x}) = {{|| \pi ( \mathbf{v}_j(\mathbf{x})   ) - \mathbf{u}_j  ||}}^{2},
\end{equation}
where $\mathbf{u}_j$ is the position of a detected landmark and $ \pi ( \mathbf{v}_j(\mathbf{x}) )$ is the corresponding mesh vertex projected into the camera space. The two terms $D_i$ and $L_j$ are combined to provide the fitting objective function:
\begin{equation}
\underset{ \mathbf{x} }  {\text{min}} \  w_d \sum_i D_i ( \mathbf{x} ) + w_l \sum_j L_j ( \mathbf{x} ) + w_{r} || \mathbf{x} ||_1,
\end{equation}
where $w_d$, $w_l$, and $w_r$ represent weights for the depth losses $D_i ( \mathbf{x} ) $, the landmark losses $L_j ( \mathbf{x} )$, and an $\mathcal{L}_1$ regularization loss, respectively. The $\mathcal{L}_1$  loss ensures a sparse solution. The minimization is carried out using a solver based on the Gauss-Seidel method.

The blendshape coefficients extracted from the videos serve as the ground-truth coefficients for training our models.

\section{Audio-visual Performance Re-targeting} \label{sec:avft}

\begin{figure}
\centering
\includegraphics[scale=0.4]{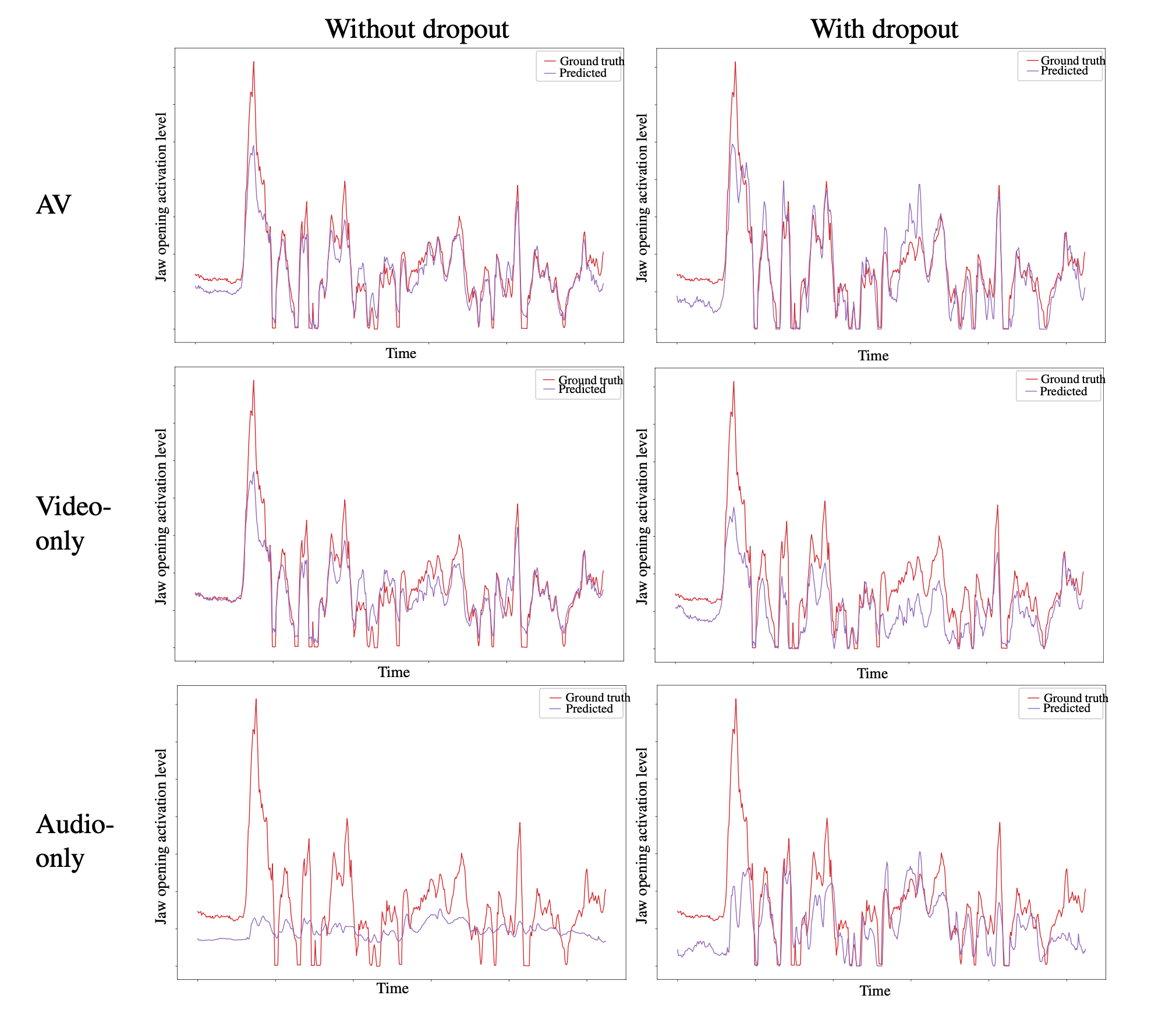}
\caption{The ground-truth (red) and predicted (blue) blendshape control for jaw opening over a sequence of speech.  The predicted trajectory was generated using an audio-visual network without modality dropout (left column) and with modality dropout (right column).  Shown are the predictions using the full audiovisual input features (top row), only the video features (middle row) and only the audio features (bottom row).  Note that the audio-only predictions without modality dropout follow the trend of the ground-truth, but the predicted blendshape has significantly less energy than is required to represent the ground-truth.  Training with modality dropout goes a long way to mitigate this limitation.}
\label{fig:av_curves}
\end{figure}

Figure~\ref{fig:framework} gives an overview of the audio-visual network for facial tracking and animation. The input visual features $f_v$ are 128x128 grey-scale images containing face crops. The face bounding boxes are  detected in each image using a pre-trained neural network.  The input acoustic features $f_a$ are 40x21 tempo-spectral features extracted from the raw speech samples. We use 40-dimensional Mel-scaled filterbank (MFB) features.  We have experimented with a both non-causal context window of size 21 that is composed of 10 past MFB frames, the current MFB frame, and 10 future MFB frames.  For real-time processing, we use a causal context window of 11 MFB, in which we eliminate the future frames.

\begin{table}[t!]
	\centering{
		\caption{The architecture of the visual feature extraction network.}
		\label{table:AVFE}
			\begin{tabular}{cccccc}
				\hline
				Type 		&  \#Filters & Kernel & Stride/padd.  & 	Output  	& 		Activation \\
				\hline                                                                   
				Conv.  		&   64		& 3x3    &	2x2/valid    & 		63x63x64	&    	RELU \\
				Conv.    	&	128		& 3x3    &	2x2/valid    & 		31x31x128	&    	RELU \\
				Conv.    	&	128		& 3x3    &	2x2/valid    & 		15x15x128	&    	RELU \\
				Conv.    	&	256		& 3x3    &	2x2/valid    & 		7x7x256		&    	RELU \\
				Conv.    	&	256		& 3x3    &	2x2/valid    & 		3x3x256		&    	RELU \\
				Conv.    	&	256		& 3x3    &	1x1/valid    & 		1x1x256		&    	RELU \\
			\hline                                                           
		\end{tabular}
	}
\end{table}

Although the network is trained end-to-end, it can be split into three modules based on their functions: 1)~a module for feature extraction, 2)~a module for multimodal fusion, and 3)~a module for regression.  In the feature extraction module, high level abstracted features are extracted from the audio and visual inputs using two stacks of convolution layers.  The first convolutional stack is applied to the images to extract 256-dimensional video embeddings $E_v$ that encode all facial expressions, including those related to speech and to head pose. The details of the convolutional layers are shown in Table \ref{table:AVFE}.
 
\begin{table}[t!]
	\centering{
		\caption{The architecture of the non-causal acoustic feature extraction network.}
		\label{table:AFEN}
			\begin{tabular}{cccccc}
				\hline
				Type 		&  \#Filters & Kernel & Stride/padd.  & 	Output  	& 		Activation \\
				\hline                                                                   
				Conv.  		&   32		& 3x3    &	2x2/valid    & 		10x19x32	&    	RELU \\
				Conv.    	&	64		& 3x3    &	1x1/valid    & 		8x17x64		&    	RELU \\
				Conv.    	&	64		& 3x3    &	1x1/valid    & 		6x15x64		&    	RELU \\
				Conv.    	&	64		& 3x3    &	1x1/valid    & 		4x13x64		&    	RELU \\
				\hline
				$\sim$ 		&  \multicolumn{3}{c}{ \#neurons }  &   	$\sim$  	& 		$\sim$  \\
				\hline
				Dense    	&	\multicolumn{3}{c}{ 256}		 		 				& 		1x1x256		&    	NONE \\
			\hline                                                           
		\end{tabular}
	}
\end{table}

The input window of acoustic MFB features are treated as an image and is pre-processed using a stack of convolutional layers to extract 256-dimensional audio embeddings $E_a$. The audio embedding captures the correlations between the pronunciation of phones and the corresponding articulator shapes.  More details about the convolutional layers for the non-causal acoustic features are shown in Table \ref{table:AFEN}.  The network architecture is almost the same for the causal acoustic features, but the stride for the first convolutional layer is $1\times1$ instead of the $2\times2$.  

As shown in Figure \ref{fig:framework}, we fuse the audio and video embeddings using concatenation, before regressing to the blendshape targets.  Note that the concatenated multimodal features are used to estimate only a subset of the blendshape weights, which includes the lips, jaw, mouth, and cheek controls ($\mathbf{x_{av}}$ in Figure \ref{fig:framework}).  The remaining blendshape targets, including the controls for the eyes and eyebrows ($\mathbf{x_{v}}$ in Figure \ref{fig:framework}), are estimated solely from the visual embeddings.

We have also examined other multimodal fusion schemes, such as adding more hidden non-linear layers after concatenation, adding canonical correlation analysis (CCA) loss \cite{thompson2005canonical} to capture the correlation between the audio and visual embeddings, gated-fusion \cite{Ovalle2017GatedMU} to adaptively estimate stream weights that capture the importance of each stream at each time-frame, and tensor-fusion \cite{Amir17} to capture the intra- and inter-modality dynamics.  However, this simple late fusion scheme with a single affine layer achieved better results than all other methods, and it is the most computationally efficient.

A challenge training a network architecture such as that described in Figure \ref{fig:framework} is the strong correlation between the input video and the output animation, which is expected since we are transferring motion from an actor to the equivalent motion on the model.  This means that although the acoustic modality adds subtlety to the output animations, most of the contribution is provided by the visual modality.  For example, the top row of Figure \ref{fig:av_curves} shows the animation control for jaw opening from an audiovisual model when driven by audiovisual inputs (left), visual-only inputs (where the audio is zeroed out) (center) and audio-only outputs (where the video is zeroed out) (right).  Notice that the video-only input does a good job of reconstructing the jaw opening control.  The corresponding control for the audio-only input has very little variation, but does tend to follow the trajectory of the ground-truth.  The combination of the audio and video (left) results in better lip-closures, which is an important cue for speech perception.  We hypothesize that encouraging the network to better exploit the acoustic information will result in generation of more accurate animation curves.

To encourage the network to pay attention to the audio inputs to further exploit the correlation between acoustic speech and lip motion, we use modality dropout \cite{Neverova2015}, where during training, batches are generated that contain either audiovisual data, video-only data (the audio is zeroed out with a given probability) and audio-only data (the video is zeroed out with a given probability).  The probabilities for dropping out the audio and video can be tuned to define how much attention the network should pay to each respective modality. The bottom row of Figure \ref{fig:av_curves} shows that the modality drop-out strategy increases the contribution of the audio stream and hence, the improves the subtlety of articulation.  The animation curves in Figure \ref{fig:av_curves} represent jaw opening, and when the value of the coefficient is zero the lips should be closed during speech.  Notice that there are several points of closure as defined in the ground-truth sequence, but the predicted animation curve for the audiovisual inputs (top row, first column) misses most of them.  The equivalent network trained using modality dropout and using audiovisual inputs hits every lip closure.  This is an important speech cue, and this subtlety has a striking effect on the quality of the animation, as discussed in Section \ref{sec:dataset}.

There are various sources of information in the inputs that we want the network to learn. Therefore, there are a number of loss terms used in the optimization.  These losses include:  1) the mean square error (MSE) loss between the ground-truth speech-related blendshape targets and the corresponding network output, 2) the MSE loss between the temporal difference for two adjacent ground-truth frames and the corresponding temporal difference in the network outputs, 3) the absolute MSE loss between the ground-truth non-speech-related blendshape controls and their corresponding network output, 4) the temporal MSE loss between the ground-truth non-speech-related blendshape controls and their corresponding network output, 5) MSE loss between the ground-truth head pose and the corresponding network predicted pose, and 6) MSE loss between the ground-truth facial landmarks and the corresponding network output.  To train with modality dropout, whenever the video input is dropped, losses 3--6 are set to zero. Since we use temporal losses to reduce jitter effects, temporally consecutive inputs should be employed or dropped out jointly. Failing to zero-out the video-only losses or to jointly dropout consecutive inputs results in poor quality output in the form of unexpected temporally adjacent frames. 

\section{Experiments and Results} \label{sec:exp}

\subsection{Dataset}\label{sec:dataset}

We recorded a dataset of ninety hours of multimodal data to train and test the audiovisual network shown in Figure \ref{fig:framework}.  The audio for each utterance in the corpus was recorded at 16kHz, 16bps PCM audio, and the video was captured at 60 frame per second (fps) RGB with corresponding 30 fps depth. The corpus contains 6847 subjects that are demographically balanced. 

Around 50,000 frames were randomly chosen from different utterances to be used as an evaluation set, and 50 complete utterances from talkers that were not used in training were held out for the subjective assessments to determine how well our approach generalizes across talkers. The remaining frames were used for training. 

\subsection{Evaluation setup}\label{sec:eval}

We use stochastic gradient descent (SGD) with Adam optimizer and a learning rate of $0.0001$, a batch size of 32, and we trained for 1~million iterations. The absolute blendshape and head pose rotation and translation loss weights are $1.0$, 1e-5, and 1e-5, respectively. The  temporal blendshape and head pose rotation and translations weights are $20$, $10$, and $0.001$, respectively. Finally, the landmark loss weight is $5.0$.

We compared six variants of our audiovisual network architecture to a baseline video-only performance-driven animation network. Four of the audiovisual networks use future and past audio context.  Of these, one system is trained without modality dropout, while the rest are trained with modality dropout. The audio dropout probabilities employed are 0.25, 0.4, and 0.5, where the video dropout probability is kept constant at 0.5 --- these values were determined empirically.  The final audiovisual network deploys only past audio context and it is trained with audio and video dropout probabilities 0.4 and 0.5, respectively. 

One of the most difficult tasks when it comes to machine learning problems involving synthesis is how to best compare models. Objective measures, such as the loss functions that the network is trained to optimize, do not usually reflect the naturalness and the quality of the network output. Human ratings of subjective quality are usually more reliable.  In this study, we use AB subjective tests to evaluate the performance of the trained networks.  In each subjective test, human graders are presented with a pair of videos and they are asked ``which video matches the speech more naturally''?   In total, 50 utterances were selected.  34 videos have good acoustic and visual conditions.  The remaining 16 videos have challenging visual or acoustic conditions, e.g.\ facial hair, glasses, difficult pose angles, or noisy acoustic conditions, to represent in-the-wild data. In all subjective tests, we use videos generated by the video-only network as a reference to compare against the videos generated by an audiovisual network.  The videos are rendered with a neutral head pose so that the graders focus only on the lip movements of the speaking face.  To prevent display ordering effects, the order in the pair that the videos are presented is randomized.  In total, 30 graders evaluated the 50 videos.  The graders are gender balanced (15 male and 15 female) and they all are native US English speakers in the age range 21--50.  

\subsection{Results}\label{sec:results}

\begin{figure}[t!]
\centering
\vspace{3mm}
\includegraphics[width=\linewidth]{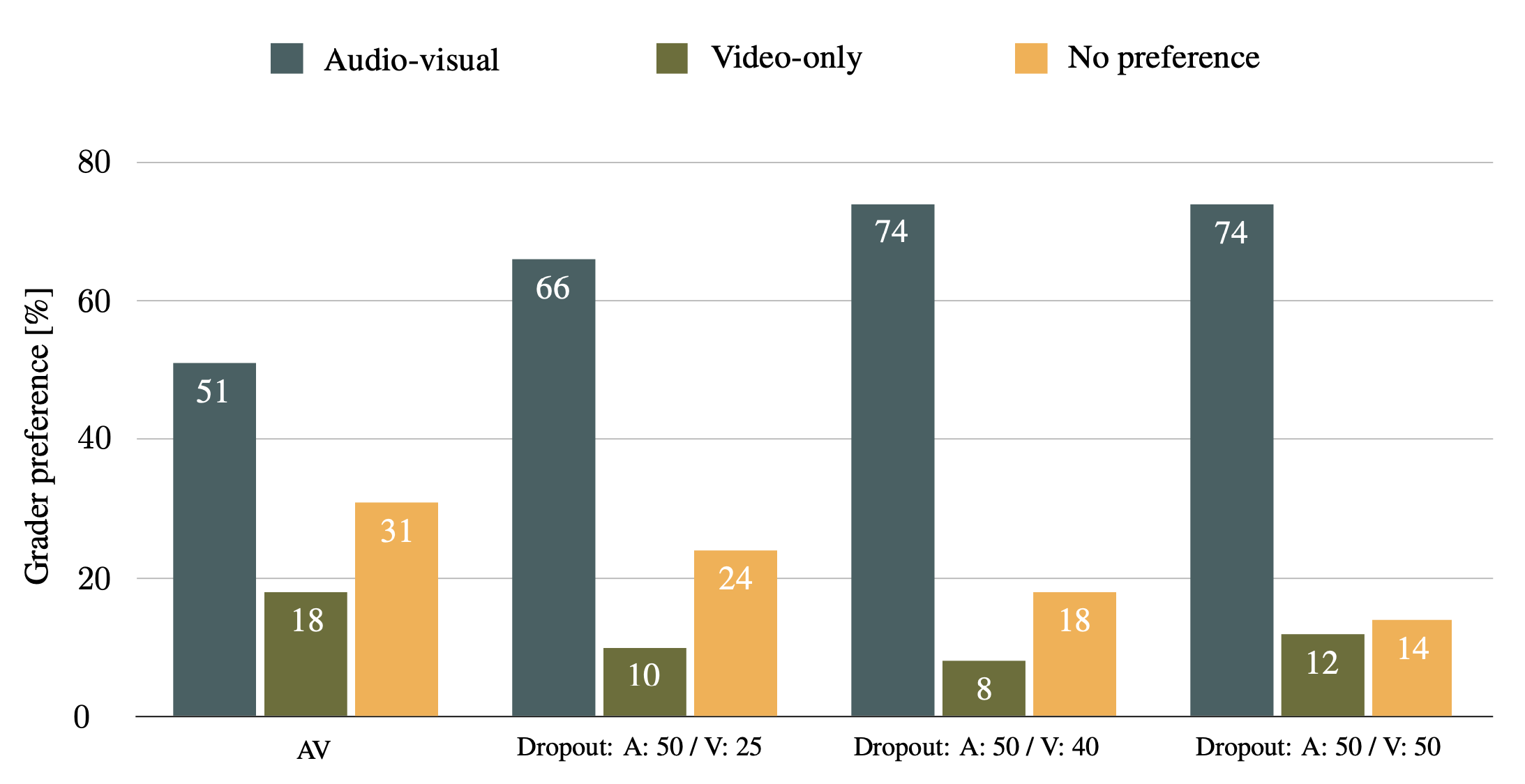}
\caption{Subjective test results comparing the performance of various audiovisual systems to the video-only system. }
\label{fig:test2}
\end{figure}

The results from four subjective tests comparing the performance of different audiovisual systems to the video-only baseline system are shown in Figure \ref{fig:test2}.   All audiovisual systems outperform the video-only system. However, the contribution of the acoustic and visual modalities is better balanced after adding modality dropout, which results in better inferred blendshape coefficients for perceptually important cues such as mouth closures.  Figure \ref{fig:test2} shows that increasing the audio-only samples in a batch increases the weight of the audio modality and thus results in the better overall performance of the audiovisual network.  In the extreme case of audio and video dropout probabilities of 0.5, where the fusion layer is not exposed to any audiovisual examples during training, the talking head sometimes over-articulates speech. However, graders still preferred the audiovisual system over the baseline.  In this instance we hypothesize that the network has not learned to fuse audio and visual information, but rather components of the network learn to focus on acoustic information and components of the network learn to focus on visual information.  The reason for the over-articulation for audiovisual inputs is because the neurons associated with both the acoustic and the visual components of the network activate simultaneously, which is not observed during training.  We are introspecting the networks to confirm that this is the case.

\begin{figure}[t!]
\centering
\vspace{3mm}
\includegraphics[width=0.6\linewidth]{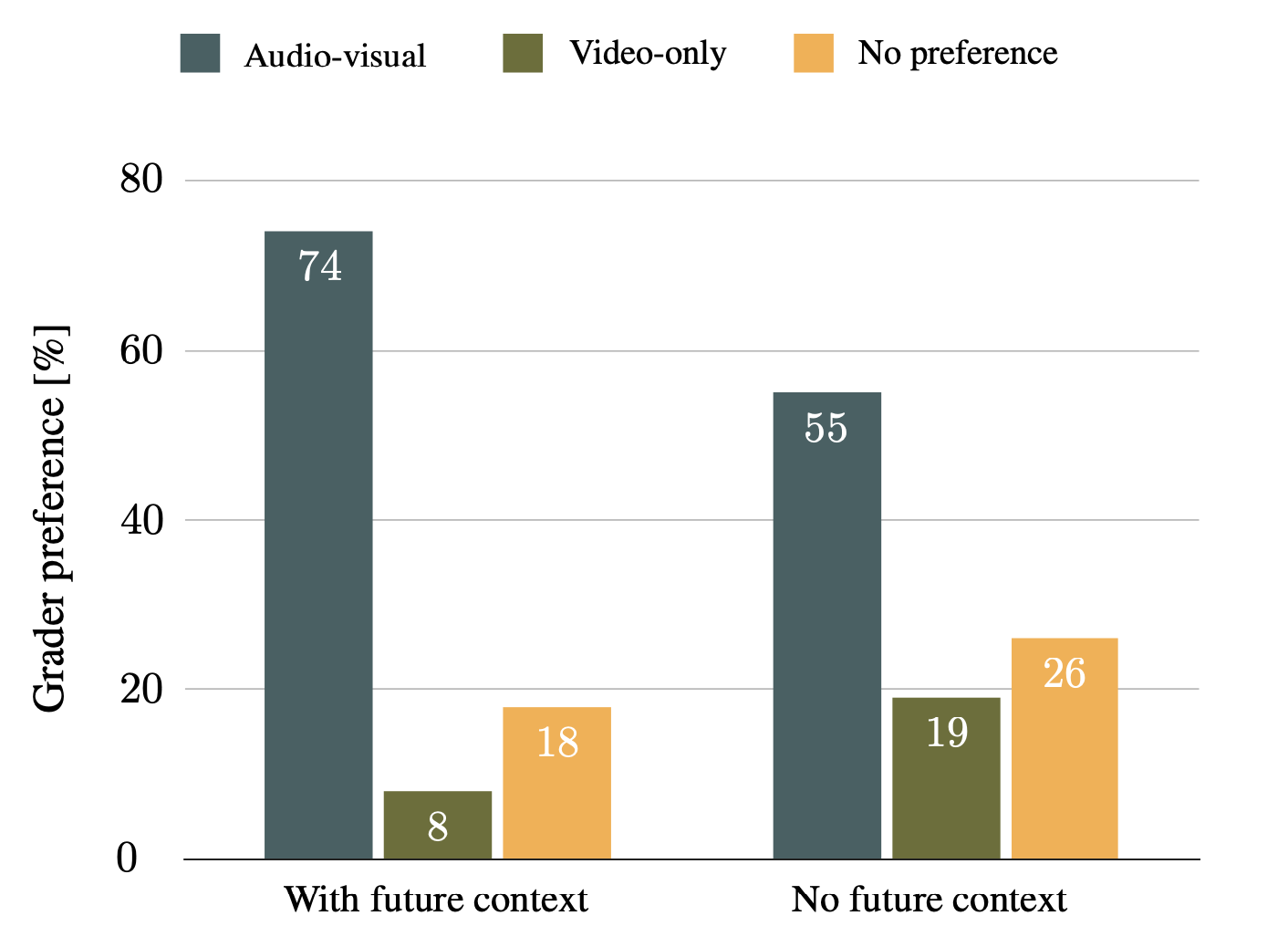}
\caption{Subjective tests comparing the baseline video-only system to two audiovisual systems that deploy audio features with and without future context.}
\label{fig:sub2_context}
\end{figure}

Figure \ref{fig:sub2_context} shows the results of the subjective tests comparing the two audiovisual systems that generated the curves in Figure \ref{fig:av_curves} to the baseline video-only system. As shown, removing the future context leads to a degradation in the audio-visual performance, which is likely because almost half of the audio features have been removed.  However, another fundamental reason could be the natural asynchrony between the audio and visual speech, where the articulators move first before speech is uttered \cite{abdelazizCHMM}.  This effect may have been compensated for by simply including future audio frames and hence, explicit modeling of the audiovisual asynchrony may not be needed.  A similar effect was  observed in audiovisual speech recognition \cite{abdelaziz2017comparing}.  To verify this we are testing the effect of increasing the number of frames when only the past context is considered.  However, from a practical implementation point of view, this would ultimately increase latency before animation can be generated and would impact a real-time system.

Balancing the trade off between training a real-time audiovisual network and the gain achieved by deploying the full context depends on the system requirements.  A semi-causal system could be used, where future audio context is restricted by the real time constraints. 
 
\subsection{Discussion}\label{sec:disc}

\begin{figure}[t!]
\centering
\vspace{3mm}
\includegraphics[width=\linewidth]{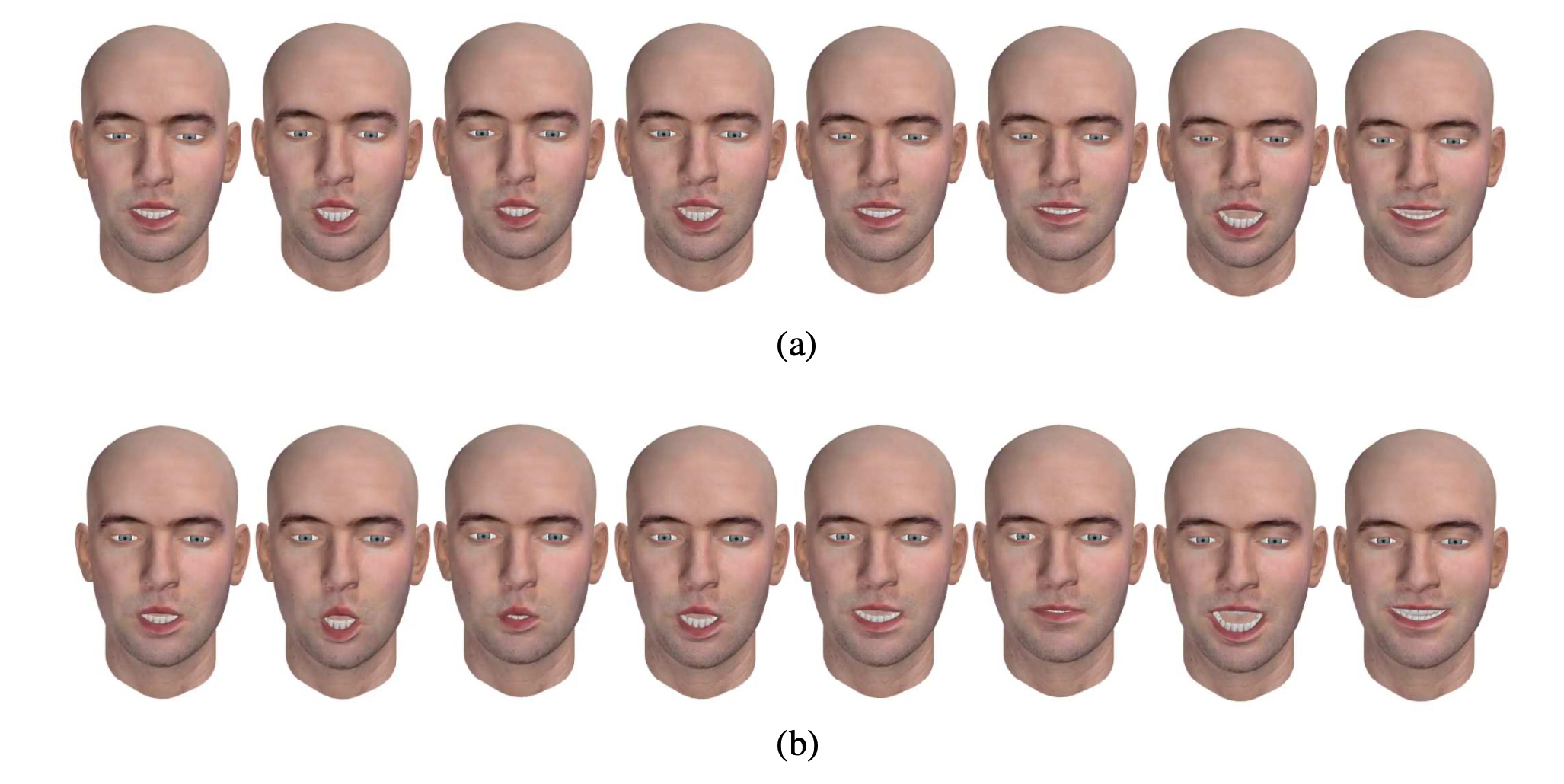}
\caption{Example frames from a video-only (a) and audiovisual (b) model speaking ``\textit{No where fast}''.  Notice the improvement in lip-rounding for ``o'' in ``no'' and ``w'' in ``where'' in frames two and three respectively, and  better representation of the f-tuck in frame six.}
\label{fig:nowherefast_n}
\end{figure}

An observation that graders frequently highlighted was that the audiovisual systems produce better articulation and agility of the lips than the video-only system. To demonstrate this, Figure \ref{fig:nowherefast_n} shows sample images from a video of a speaker saying ``No where fast''.   The audiovisual system in Figure \ref{fig:nowherefast_n}-(b) gives better rounding of the lips for the letter ``o'' in the second frame and ``w'' in the third frame. It also closes the mouth more accurately for the letter ``f'' in the sixth frame.  

An aside learned by the audiovisual system is that speakers can not generate speech sounds when their lips are closed.  This is helpful in avoiding animating the lips to match background speech or acoustic noise.  Unless the signal-to-nose ratio (SNR) is very low, the lip motion of the avatar is barely affected by any background speech or acoustic noise. 

\section{Conclusion} \label{sec:conclusion}
In this paper, we have presented a neural network-based approach for driving 3D talking faces using audiovisual data. The neural network extracts audio embeddings from audio spectral features and visual embeddings from face images. The audio and visual embeddings are fused using an affine layer and regressed to speech-related facial controls. Non-speech-related facial controls and head pose are inferred from only the video embeddings. For training our audiovisual network, we have used modality dropout, where audio or visual features are dropped in each batch according to pre-defined dropout probabilities.  This strategy increases the contribution of the audio features to the overall performance, and we have shown using subjective assessment that this in turn enhances the lip retargeting performance.  Finally, we have shown that removing the future context without increasing the number of past frames impacts the quality of the animation.  We are investigating semi-causal audio features to balance between real-time constraints and reducing the impact on the quality of the animation.

\section{Acknowledgments} \label{sec:acksn}
The authors would like to express their appreciation to Russ Webb, Saurabh Adya, Ashish Shrivastava, and Thibault Weiss for their many useful inputs and valuable comments.

\bibliographystyle{plain}  
\bibliography{bibliography} 

\end{document}